\documentclass{iopconfser}
\usepackage[utf8]{inputenc}

\usepackage{xcolor}
\usepackage{graphicx}
\usepackage{hyperref}
\usepackage{amsmath}
\usepackage{amssymb}
\usepackage{xspace}

\newcommand{\AthenaK}{\texttt{AthenaK}\xspace}

\begin{document}

\title{AthenaK Simulations of Magnetized Binary Neutron Star Mergers}

\author{Jacob Fields$^{1}$ and David Radice$^{2,3,4}$}

\affil{
$^1$School of Natural Sciences, Institute for Advanced Study, Princeton, USA\\
$^2$Institute for Gravitation and the Cosmos, The Pennsylvania State University, University Park, USA \\
$^3$Department of Physics, The Pennsylvania State University, University Park, USA\\
$^4$Department of Astronomy \& Astrophysics, The Pennsylvania State University, University Park, USA}

\email{fields@ias.edu}

\begin{abstract}
We present new numerical-relativity simulations of a magnetized binary neutron
star merger performed with \AthenaK. The simulations employ a temperature- and
composition-dependent tabulated nuclear equation of state, with initially dipolar
fields with a maximum initial strength of ${\sim}10^{16}\ {\rm G}$
which extend outside the stars. We employ adaptive mesh refinement and
consider three grid resolutions, with grid spacing down to $\Delta x_{\rm min}
\simeq 92\ {\rm m}$ in the most refined region. When comparing the two highest
resolution simulations, we find orbital dephasing of over 7 orbits until merger
of only $0.06$ radians. The magnetic field is amplified during the merger and we
observe the formation of a magnetized funnel in the polar region of the remnant.
Simulations are continued until about $30$ milliseconds after merger. However,
due to significant baryonic pollution, the binary fails to produce a
magnetically-dominated outflow. Finally, we discuss possible numerical and
physical effects that might alter this outcome.
\end{abstract}

\section{Introduction}
Binary neutron star (BNS) mergers are valuable events for multi-messenger astronomy, as they produce
gravitational waves observable to ground-based detectors like LIGO and Virgo \cite{LIGOScientific:2017vwq} and
kilonovae and short gamma-ray bursts visible to electromagnetic telescopes \cite{LIGOScientific:2017ync}.
Because these events include important interactions from all four fundamental forces and occur at extragalactic
distances, they are extremely rich sources of information which probe the edges of current physics. For example,
observations of GW170817 alone placed stringent constraints on violations of general relativity \cite{LIGOScientific:2017zic},
provided new information about the nuclear equation of state \cite{Shibata:2017xdx,Ruiz:2017due,Margalit:2017dij,
Radice:2017lry,De:2018uhw}, and provided an independent measurement of the Hubble constant \cite{LIGOScientific:2017adf,
Hotokezaka:2018dfi,Dietrich:2020efo}.

Interpreting these merger events requires accurate models. Through the late inspiral and post-merger phases,
the only ab-initio tool which can capture the interaction between strong gravity, nuclear matter, radiation,
and large magnetic fields is numerical relativity. Such models require significant computational resources, particularly
as our observational capabilities improve. This creates a persistent need for better numerical relativity tools, such as
employing more accurate numerical methods \cite{Radice:2013hxh,Most:2019kfe,Doulis:2022vkx,Kiuchi:2022ubj,
Deppe:2024ckt,Adhikari:2025nio,Kiuchi:2025ksk}, building codes which are better suited to modern large-scale computational
resources \cite{Kidder:2016hev,Cook:2023bag,Shankar:2022ful,Kalinani:2024rbk,Fields:2024pob,Palenzuela:2025ucx}, and
including more realistic physics \cite{Ruffert:1995fs, Rosswog:2003rv, Sekiguchi:2011zd, Sekiguchi:2015dma, Foucart:2015vpa, Foucart:2016rxm, Radice:2017zta, Miller:2019dpt, Foucart:2020qjb, Radice:2021jtw}.

In this work, we demonstrate some of the capabilities of the GPU-accelerated \AthenaK astrophysics code by modeling a BNS
system with a finite-temperature nuclear equation of state with dipole magnetic fields. Sec.~\ref{sec:methods} details our
numerical setup and initial data. In Sec.~\ref{sec:results}, we discuss the accuracy of our gravitational waveforms and discuss
qualitative aspects of the post-merger dynamics, and we briefly highlight the computational performance of \AthenaK. Finally, we
conclude in Sec.~\ref{sec:conclusion} and highlight potential avenues for future work.

\section{Methods}
\label{sec:methods}
The initial data consists of an equal mass quasicircular binary neutron star system with $45~\mathrm{km}$
initial separation. Each star has a gravitational mass of $1.3~\mathrm{M}_\odot$ and is modeled using the
SFHo \cite{MOLLER1997131,Hempel:2009mc,Hempel:2011mk,Steiner:2012rk} equation of state (EOS) and assumed to be
in cold beta equilibrium. We construct the initial data using the LORENE pseudospectral code
\cite{PhysRevD.63.064029,2016ascl.soft08018G}.

We evolve the system using the \AthenaK astrophysics code \cite{2024arXiv240916053S,Zhu:2024utz,Fields:2024pob}
using the HLLE approximate Riemann solver \cite{Harten:1983,EINFELDT1991273} with fifth-order WENOZ
\cite{BORGES20083191} reconstruction. To reduce the frequency of flooring and improve robustness, we also employ a
first-order flux correction \cite{Lemaster:2008gh}. The divergence-free condition for the magnetic field is maintained via
an upwind constrained transport scheme \cite{Gardiner:2005hy,Gardiner:2007nc}. We modify the SFHo EOS to include
the contributions of trapped neutrinos using the prescription in Ref.~\cite{Perego:2019adq}, which allows us to
approximate the effect of neutrinos in the optically thick regime by advecting the lepton fraction \cite{Espino:2023dei}. To improve performance,
we also retabulate the EOS so that the density and temperature are spaced evenly in ``not-quite-transcendental'' function
space rather than true logarithmic space \cite{Hammond:2025gjd}.

The evolution grid is a Cartesian box spanning $[-1536~G\mathrm{M}_\odot/c^2,1536~G\mathrm{M}_\odot/c^2]\approx
[-2268~\mathrm{km},2268~\mathrm{km}]$ in all three directions. The base grid contains $192$, $384$, or $768$ cells
for the low-resolution (LR), medium-resolution (MR), or high-resolution (HR) run, respectively. We apply six
levels of adaptive mesh refinement (AMR), with the finest level covering a $10~G\mathrm{M}_\odot/c^2\approx
14.8~\mathrm{km}$ radius around each star for an effective resolution of ${\sim}369~\mathrm{m}$, ${\sim}185~\mathrm{m}$,
or ${\sim}92~\mathrm{m}$ in these regions. The AMR tracks the location of the stars by following the local minimum of the
lapse function $\alpha$.

We superimpose a dipolar magnetic field on top of each neutron star. To ensure that the initial discretized field
satisfies $\nabla\cdot\mathbf{B}=0$ to machine precision, we compute the magnetic field from the vector potential
modeled at cell edges with
\begin{equation}
   A_\phi = \frac{4r_0^3 B_0}{23(r_0^2 + r^2)^{3/2}}
     \left(1 + \frac{15r_0^2}{8}\frac{r_0^2 + x^2 + y^2}{\left(r_0^2 + r^2\right)^2}\right),
\end{equation}
where $r_0$ is the position of the current generating the field and $B_0$ is the maximum magnetic field\footnote{Note that this
fixes the \textit{densitized} magnetic field $\sqrt{\gamma}\mathbf{B}$, where $\gamma$ is the determinant of the metric. In practice
$\max|B|$ is about a factor of two smaller than $B_0$ would suggest.}. This field
extends past the surface of the stars, causing the artificial atmosphere (set to
$\rho_\mathrm{atm}\approx1.85\times10^3~\mathrm{g}/\mathrm{cm}^3$) to be strongly magnetized, especially near the surface
of the stars. We choose $r_0 = 2.13~G\mathrm{M}_\odot/c^2\approx3.14~\mathrm{km}$ and $B_0=10^{16}~\mathrm{G}$. Though this
field is much larger than the $\lesssim10^{13}~\mathrm{G}$ fields expected in a realistic inspiral, the field is expected to be rapidly
amplified up to ${\sim}10^{16}~\mathrm{G}$ in the early post-merger phase. Since the length scales needed to resolve this
amplification are too short to resolve with current computational capabilities, it is standard practice to impose a field
much closer to the expected post-merger field during the inspiral \cite{Price:2006fi, Kiuchi:2015sga, Chabanov:2022twz, Kiuchi:2023obe, Aguilera-Miret:2025nts}. We do not impose reflection symmetry across the $z=0$ plane \cite{Gutierrez:2025gkx, Cook:2025frw}. Each run evolves for $10^4~ G \mathrm{M}_\odot/c^3\approx49~\mathrm{ms}$.

\section{Results}
\label{sec:results}
\begin{figure}
    \centering
    \includegraphics[width=\textwidth]{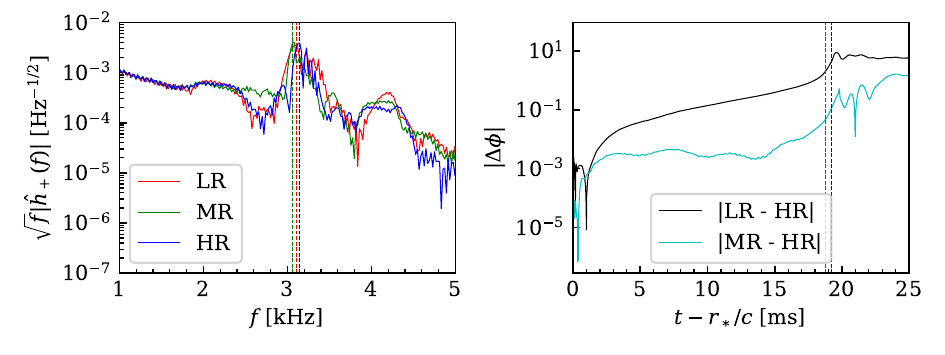}
    \caption{\label{fig:waveform} (Left) The power spectrum of the real part of the $(\ell=2,m=2)$ mode of the gravitational-wave
        strain for each resolution. The dashed vertical lines mark $f_2$, the peak frequency of the post-merger phase. (Right) The phase 
        difference of the LR and MR runs measured relative to the HR run. The dashed vertical lines mark the merger time for each run
        using the same colors as the left plot. The waveform is extracted at a radius of
        $r=400~G\mathrm{M}_\odot/c^2\approx591~\mathrm{km}$ from the origin.}
\end{figure}

The binary completes ${\sim}7$ orbits before merging and produces a long-lived
remnant massive neutron star (RMNS) \cite{Radice:2018xqa}.
These dynamics are encoded in the gravitational-wave strain, for which we show the
$(\ell=2,m=2)$-mode power spectrum and the dephasing between resolutions in Fig.~\ref{fig:waveform}.
The time of merger, defined as the time when the
amplitude of the complex waveform $|h_+ - i\, h_\times|$ peaks, slightly increases with resolution from
$t-r_\ast/c\approx18.81~\mathrm{ms}$ for LR to
$t-r_\ast/c\approx19.22~\mathrm{ms}$ for MR and
$t-r_\ast/c\approx19.24~\mathrm{ms}$ for HR. The corresponding dephasing at merger between LR and HR, and MR and HR resolutions are of
$4.5$ and $red0.12$ radians respectively. In terms of orbital dephasing, this corresponds to a difference of $0.06$ radians. The phase differences between the MR and HR simulations are flat (a few
$\times 10^{-3}$) over most of the inspiral and only start growing when the stars come into contact. This shows that \AthenaK can achieve
an accuracy comparable to, or better than, state-of-the-art ideal-fluid inspiral simulations
\cite{Bernuzzi:2012ci, Radice:2013hxh, Dietrich:2017aum, Foucart:2019yzo, Kiuchi:2019kzt, Habib:2025bkb}, even when employing full 
temperature and composition-dependent EoS and with magnetic fields. We do not see clean second-order convergence in the inspiral as
might be expected, even when using phase-aligned waveforms. The most likely cause for this is that the LR data is too coarse, so the
waveform is either not in a convergent regime or is not solely dominated by errors in the fluid evolution. The post-merger signal is
nearly monochromatic, as evidenced by the  strong peak in the power spectrum at ${\sim}3~\mathrm{kHz}$, and decays over a timescale of 
${\sim}20~{\rm ms}$ of the merger
\cite{Shibata:2005ss, Bauswein:2011tp, Takami:2014tva, Bernuzzi:2015rla, Bernuzzi:2015opx, Zappa:2017xba}. Here, we observe excellent 
agreement between the three resolutions with $\Delta f \lesssim {82}\ {\rm Hz}$, comparable to the nominal uncertainty in the 
frequency for a ${\sim}30\ {\rm ms}$ signal. This demonstrates that \AthenaK is well suited for quantitative gravitational-wave astronomy 
modeling.

\begin{figure}
    \centering
    \includegraphics[width=0.85\textwidth]{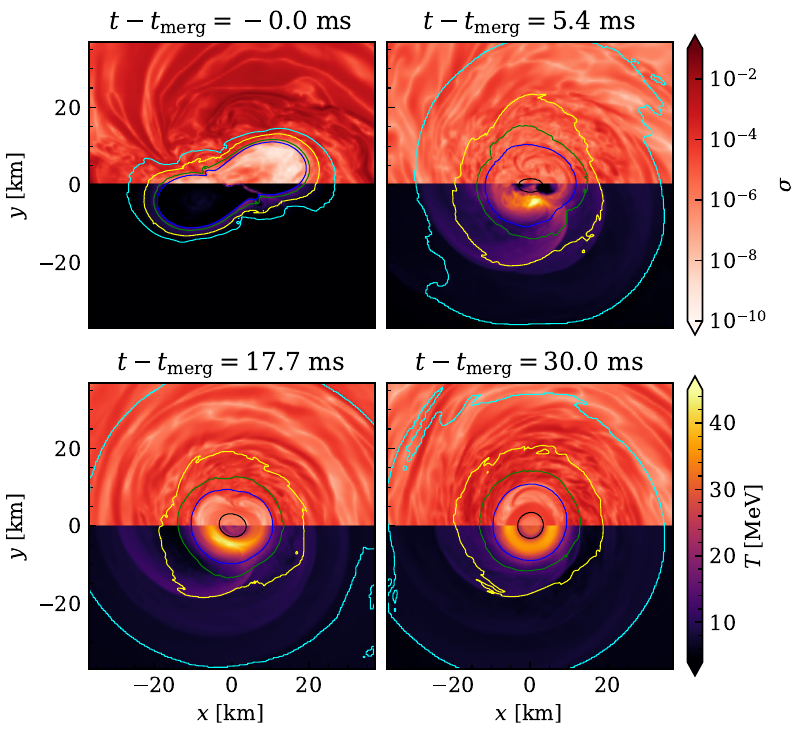}
    \caption{\label{fig:slice_plot} A slice plot of the magnetization $\sigma=b^2/\rho$ and temperature in the $xy$ plane of the HR run
        at various times relative to merger. The cyan, yellow, green, blue, and
        black contours correspond to rest-mass densities of $10^{11}$, $10^{12}$, $10^{13}$, $10^{14}$, and
        $10^{15}~\mathrm{g}/\mathrm{cm}^{3}$, respectively.}
\end{figure}

We show snapshots of the evolution for the HR binary at various times in
Fig.~\ref{fig:slice_plot}. Throughout the inspiral phase, temperatures remain
low, and the strongest magnetic fields are located in the center of each star.
During the merger, the shearing of the surfaces leads to large temperature
spikes and amplification of the magnetic field via the Kelvin-Helmholtz
instability. A high-density core with rest-mass density
$\rho>10^{15}~\mathrm{g}/\mathrm{cm}^3$ rapidly forms. As the remnant settles,
the amplified field and strongest temperatures are found in a toroidal shear
layer with densities of several times $10^{14}~\mathrm{g}/\mathrm{cm}^3$. This
is in good agreement with prior results from the literature
\cite{Giacomazzo:2014qba, Bernuzzi:2015opx, Kastaun:2016yaf, Hanauske:2016gia,
Kiuchi:2017zzg, Palenzuela:2021gdo, Radice:2023zlw, Combi:2023yav,
Gutierrez:2025gkx, Cook:2025frw, Aguilera-Miret:2025nts}. The exterior of the
stars, initially strongly magnetically dominated, is quickly polluted by baryons
lifted from the surface of the stars as result of the artificial heating of the
star surfaces \cite{Gittins:2024jui}, as a result the magnetization drops to
$\sigma = b^2/(\rho c^2) \lesssim 10^{-2}$ already early on in the inspiral. The
magnetization is further reduced after merger, when mass ejection are driven by
shocks and tidal interaction between the stars. Material squeezed from the
collisional interace between the stars forms a warm $T \sim 10\ {\rm MeV}$,
magnetized ($\sigma \sim 10^{-5}$) torus~\cite{Cook:2025frw}.

\begin{figure}
    \centering
    \includegraphics[width=1.0\textwidth]{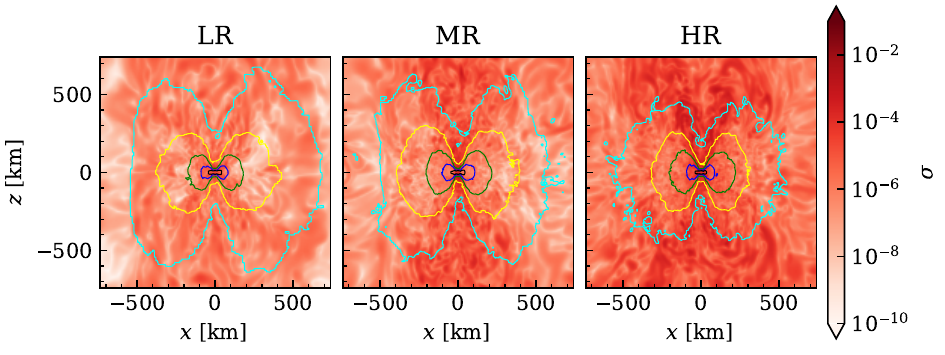}
    \caption{\label{fig:mag_plot} Slice plots of the magnetization $\sigma=b^2/\rho$ in the
    $xz$ plane for the LR, MR, and HR runs at $t-t_\mathrm{merg}\approx30~\mathrm{ms}$. The cyan, yellow, green, blue, and black
    contours correspond to rest-mass densities of $10^7$, $10^8$, $10^9$, $10^{10}$, and $10^{11}~\mathrm{g}/\mathrm{cm}^3$.}
\end{figure}

Fig.~\ref{fig:mag_plot} shows the magnetization of the remnant at
${\sim}30~\mathrm{ms}$ post-merger for all three resolutions. There is evidence
of a magnetically-dominated funnel region forming by this time, with higher
resolution simulations producing higher magnetization. However, the
magnetization values observed in the simulations are too small to produce a
relativistic outflow. Even in the HR run, $\sigma\lesssim0.1$ due to baryon
pollution, suggesting that this funnel cannot yet support the relativistic jet
needed to launch a short gamma-ray burst (GRB). It is possible that simulations
avoiding the artificial heating of the stars, for example using a better Riemann
solver \cite{Kiuchi:2022ubj, Xie:2024iih}, will show reduced baryon pollution
and might result in jet launching shortly after merger. Neutrinos might also
contribute by reducing the baryon pollution at high latitudes
\cite{Radice:2016dwd, Ciolfi:2020hgg, Mosta:2020hlh, Combi:2023yav,
Musolino:2024sju}.

The estimated cost of each run was $840$ GPU-hours (210 node-hours)  for LR,
$5800$ GPU-hours (1450 node-hours) for MR, and $65000$ GPU-hours (16250
node-hours) for HR on NERSC Perlmutter, which contains four Nvidia A100s per GPU
node. Since the resolution doubles with each run, naively one would expect the
cost to increase by a factor of $16$. However, we observe that the MR and HR
runs are considerably cheaper than this scaling would predict. This is because
\AthenaK uses block-based octree AMR, so mesh refinement occurs only along block
boundaries and must enforce a 2:1 constraint. Therefore, by keeping the
meshblock size fixed as resolution increases, the mesh structure becomes more
efficient because it is less likely to overrefine regions far away from the
star.

\section{Conclusion}
\label{sec:conclusion}
We have presented new GRMHD simulations of magnetized binary neutron star
mergers with composition and temperature-dependent nuclear equation-of-state
performed with \AthenaK. These simulations demonstrate the capabilities of our
new GPU-accelerated numerical-relativity infrastructure. \AthenaK achieves high
accuracy in the inspiral, with orbital dephasing at merger comparable to that
obtained in more specialized simulations, which trade accuracy in the inspiral
for physical realism in the postmerger by dropping magnetic fields and employing
idealized equations-of-state. With \AthenaK these compromises are not necessary.
The postmerger evolution shows the launching of magnetized outflows and the
formation of a massive, long-lived remnant, surrounded by a magnetized torus, in
lines with previous results from the literature. We observe a progressive
increase in the magnetization of the funnel region along the rotational axis of
the remnant with resolution. However, none of our simulations achieve magnetic
domination or launch relativistic jets. We speculate that the baryon pollution
in this region is a numerical artifact and arises because the surfaces of the
stars unphysically heat up during the inspiral, a well known numerical problem
in neutron star merger simulations, and that sufficiently well resolved
simulations, or simulations with better star surface preservation might launch
jets in the early postmerger. At the same time we cannot exclude that neutrinos
and/or black-hole formation will be required to successfully launch jets in the
postmerger.

The \AthenaK numerical-relativity infrastructure is still under active development. Among the new features being tested are an improved Riemann solver based on the HLLD scheme \cite{Mignone:2008ii, Kiuchi:2022ubj, Xie:2024iih}, which might mitigate the artificial heating of the stars during the inspiral, an M1 neutrino transport scheme, based on the formalism we described in \cite{Radice:2021jtw}, and a full-Boltzmann neutrino-transport scheme based on the $F\!P_N$ method \cite{Radice:2012tg, Bhattacharyya:2022bzf}.

\section*{Acknowledgments}
It is a pleasure to thank Eduardo Guti\'errez for discussions and all the
\AthenaK developers.
This work was supported by NASA under award No. 80NSSC25K7213. DR also
acknowledges support from the Sloan Foundation, the U.S. Department of Energy,
Office of Science, Division of Nuclear Physics under Award Number(s)
DE-SC0021177 and DE-SC0024388, and from the National Science Foundation under
Grants No. PHY-2407681 and PHY-2512802. 
This research used resources of the National Energy Research Scientific Computing Center (NERSC), a Department of Energy User Facility using NERSC award NP-ERCAP0031370.

\bibliographystyle{iopart-num.bst}
\bibliography{references.bib}

\end{document}